\documentclass[sigconf]{acmart}

\usepackage{xurl}
\usepackage{booktabs}
\usepackage{tabularx}
\usepackage{tablefootnote}
\usepackage{comment}
\usepackage{fancyvrb}
\usepackage{subcaption}
\usepackage{float}
\usepackage{listings}
\usepackage{makecell}
\usepackage{amsmath}
\usepackage{multirow}
\usepackage{arydshln}
\usepackage{seqsplit}
\usepackage[gen]{eurosym}
\usepackage{xcolor}

\floatstyle{plain}
\newfloat{listing}{htbp}{lol}
\floatname{listing}{Listing}
\usepackage[ruled,vlined]{algorithm2e}

\lstdefinestyle{PHP}{
  language=PHP,
  basicstyle=\ttfamily\scriptsize,
  keywordstyle=\color[HTML]{0000FF}\bfseries,       
  stringstyle=\color[HTML]{CC0000},                  
  commentstyle=\color[HTML]{008000}\itshape,         
  numbers=left,
  showstringspaces=false,
  breaklines=true,
  tabsize=4,
  alsoletter={\$},
  morekeywords=[2]{echo, print, isset, empty, count, strlen,
                   str_replace, preg_match, preg_replace,
                   explode, implode, json_encode, json_decode,
                   base64_encode, esc_attr},
  keywordstyle=[2]\color[HTML]{9900CC}\bfseries,    
  morekeywords=[3]{\$result, \$link, \$args, \$selector},
  keywordstyle=[3]\color[HTML]{007777}\bfseries,    
  escapeinside={(*}{*)},
}

\usepackage{upquote}   

\definecolor{mdGray}{HTML}{6A737D}   
\definecolor{mdBlue}{HTML}{0366D6}   
\definecolor{mdGreen}{HTML}{22863A}  
\definecolor{mdPurple}{HTML}{6F42C1} 
\definecolor{mdRed}{HTML}{D73A49}    
\definecolor{mdBg}{HTML}{F6F8FA}     

\lstdefinelanguage{Markdown}{
  sensitive=true,
  morecomment=[l][\color{mdGray}\itshape]{>},
  morecomment=[l][\color{mdPurple}\bfseries]{\#},
  moredelim=[is][\bfseries]{**}{**},
  moredelim=[is][\itshape]{_}{_},
  moredelim=[s][\color{mdGreen}\ttfamily]{`}{`},
}

\lstdefinestyle{markdown}{
  language=Markdown,
  basicstyle=\ttfamily\scriptsize,
  breaklines=true,
  breakatwhitespace=true,
  showstringspaces=false,
  upquote=true,
  rulecolor=\color{mdGray!40},
  tabsize=2,
}

\usepackage{cleveref}
\crefname{lstlisting}{listing}{listings}
\Crefname{lstlisting}{Listing}{Listings}
\crefname{algorithm}{algorithm}{algorithms}
\Crefname{algorithm}{Algorithm}{Algorithms}
\crefname{section}{section}{sections}
\Crefname{section}{Section}{Sections}
\crefname{appendix}{appendix}{appendices}
\Crefname{appendix}{Appendix}{Appendices}

\AtBeginDocument{}

\copyrightyear{2026}
\acmYear{2026}
\setcopyright{cc}
\setcctype{by}
\acmConference[AISec '26]{19th Workshop on Artificial Intelligence and Security} 
{November 15--19, 2026}{The Hague, Netherlands}
\acmBooktitle{19th Workshop on Artificial Intelligence and Security (AISec '26), 
November 15--19, 2026, The Hague, Netherlands}
\acmDOI{10.1145/3847352.3848121}
\acmISBN{979-8-4007-3031-3/2026/11}

\begin{document}

\title[Plug 'n' Pray: Agentic LLM-based Detection of Potential Log File Exposures]{Plug 'n' Pray: Agentic LLM-based Detection of Potential Log File Exposures in Third-Party Content Management System Plugins}

\author{Sebastian Neef}
\affiliation{%
  \institution{Technische Universität Berlin}
  \city{Berlin}
  \country{Germany}
}
\email{neef@sect.tu-berlin.de}
\renewcommand{\shortauthors}{S. Neef}

\begin{abstract}
Content Management Systems (CMS), such as WordPress, power a large share of the web ($\sim$58\%), and their extensibility through third-party plugins is a major source of their popularity as well as of their attack surface.
One high-impact weakness that remains understudied is log file exposure by CMS plugins, which create log files for debugging or other purposes. If these files are insufficiently secured, they can disclose sensitive information (e.g. credentials, personal data) which has led to website compromises in the past.

In this work, we present an agentic, LLM-based framework that automatically detects potential log file exposures in plugins of the most popular CMS (WordPress). Our agent analyzes each plugin by performing static and dynamic analysis. 

We evaluated our approach on the 300 most-installed WordPress plugins (about 0.6\% of all), which together account for over 250M active installations, i.e. 75\% of all active installations in the official plugin ecosystem. We manually validated each finding, reproducing 79 of 81 findings from 62 plugins.
We observed that several protective measures appear to be implemented that we classify as \emph{creation-control} (e.g. manual log activation) and \emph{access-control} (e.g. deny rules in \texttt{.htaccess}). However, we find that multi-layered protection is required, but not always present. 

From these results we derive a taxonomy of log file path and protection patterns and deduce a set of best practices for developers to securely handle them. Finally, our study corroborates that agentic LLMs are an useful tool for security analysis.
\end{abstract}



\maketitle

\section{Introduction}\label{sec:introduction}
The internet has become a crucial part for the lives of almost 75\% of the world's population \cite{itu2025internetusers}, e.g. allowing them to exchange information online or do online-shopping. 
Since creating websites can be a complex task, Content Management Systems (CMS) aim to reduce that barrier to entry by offering ready-to-use and extensible frameworks that allow the user to focus on the content rather than technical aspects. 
For example, the most popular CMS is WordPress\footnote{\url{https://wordpress.org/}, accessed: 2026-09-15} which covers almost 60\% of all CMS-based websites and over 41.5\% of all websites according to W3Techs \cite{w3techs2026cms}.
One plausible reason for the popularity  of CMS' is the feature-richness and extensibility achieved through plugin systems, which allow users to install third-party developed extensions. 

Although plugin submissions undergo review for popular CMS \cite{wordpressYourPlugin,joomlaSubmittingExtension,wixAboutDistribution}, many security issues are being discovered in such plugins \cite{wpscan2026statistics,joomlaVulnerableExtensions}. 
One potentially severe security issue arises when plugins create and fail to protect log files with sensitive content from being accessed by unauthorized actors, which could lead to the compromise of the website.
For example, the easy-wp-smtp plugin leaked the admin account's password reset token in a log file \cite{easywpsmtp0day}, and many other plugins were found to have exposed log files in an industry study in 2020 \cite{detectifyWordpressPlugins}.

Recently, Large-Language Models (LLMs) have shown to be a promising tool to identify security issues and vulnerabilities in software \cite{anthropicProjectGlasswing,LU2024112031,guo2024outside,du2024vul,cao2024realvul,fang2024llmagentsautonomouslyexploit}, however their application to the identification of log file exposure remains yet to be explored in an academic setting.

Thus, this work extends the body of literature with the following contributions:
\begin{itemize}
    \item An agentic LLM-based log file exposure detection framework that answers \textbf{RQ1}: \emph{How prevalent are log file exposures in WordPress plugins?}
    \item An extended taxonomy of common patterns for log file exposures based on the answer to \textbf{RQ2}: \emph{What log file paths and protection mechanisms patterns can be identified in popular WordPress plugins?}
    \item Best practices for developers derived from the previous results answering \textbf{RQ3}: \emph{How to securely create log files in WordPress and other CMS plugins?}
\end{itemize}

\section{Background and Related Work}\label{sec:backgroundrelwork}
This section will provide the reader with the necessary background on this topic and bring this work into the context of the existing body of literature.

\subsection{Content Management Systems and Third-Party Plugins}
Many different Content Management Systems (CMS) exist to help non-technical users create their own website by abstracting away from technical aspects.
Some CMS are operated and hosted by corporate entities and available on a subscription basis, e.g. Shopify\footnote{\url{https://www.shopify.com/}, accessed: 2026-09-15}, Wix\footnote{\url{https://www.wix.com/}, accessed: 2026-09-15}, while others are open-source and can be self-hosted, e.g. WordPress, Joomla\footnote{\url{https://www.joomla.org/}, accessed: 2026-09-15}.
Thus, CMS are a cost-effective way for a wide range of users, from beginners to professionals, to develop websites. 
According to W3Techs, CMS are widely adopted \cite{w3techs2026cms}: The top 10 CMS cover over 58\% of all internet websites, with WordPress, Shopify, Wix, Squarespace, and Joomla being the most popular ones. 
With a market share of about 60\%, the open-source CMS WordPress is over 50\% ahead of its competitors.

CMS can be specialized on specific use-cases, e.g. e-commerce, blogs, or website building, but often their plugin and extension systems allow users to extend the features or functionality by installing third-party plugins.
For example, Woocommerce\footnote{\url{https://woocommerce.com/}, accessed: 2026-09-15} turns the WordPress blog system into an online-shop, or Elementor\footnote{\url{https://wordpress.org/plugins/elementor/}, accessed: 2026-09-15} into a drag-and-drop website builder. 
In fact, the landscape of plugins is huge: The official WordPress plugin repository counts over 60,000 plugins with several reaching over 10M active installations \cite{wpodysseyManyWordPress}.

However, since anyone can develop and submit their own plugin to the respective plugin stores, the code quality and security of such plugins may vary, despite submission reviews.
Several plugins have evolved into depending on other plugins for their functionality, which can complicate the plugin update process \cite{coevolution}.
Unsurprisingly, a dedicated project emerged to track vulnerabilities in WordPress and its plugins or themes: 
The WPScan vulnerability database contains over 73,000 reported vulnerabilities \cite{wpscan2026statistics}. 

With its openness, popularity, and rich plugin ecosystem, WordPress has been used and studied in prior academic work. 
Singh found WordPress to be the most accessible CMS for academic staff, compared to Drupal and Joomla \cite{singhcomparative}.
However, the low barrier to entry might lead to users not maintaining their WordPress instance correctly. 
A study from 2023 found that WordPress websites are not always up to date and can lack important security patches \cite{ekstam2023vulnerabilities}.
Although vulnerabilities in WordPress itself exist, installed plugins impose a larger attack surface \cite{kabata2025analysis}.
In 2025, 91\% of the vulnerabilities were found in plugins, 9\% in themes and only 6 issues in WordPress \cite{patchstackStateWordPress}.
In fact, installing plugins or blindly updating them is a security risk, as plugins can become malicious or contain malware, as shown by Kasturi et al. \cite{mistrust}.
According to a study by Koskinen et al. in 2012, there is no clear correlation between the security of a plugin and its rating \cite{qualityofWPplugins}. 
In 2019, however, Ruohonen found that from a demand-side viewpoint, a higher install count is statistically associated with a higher number of vulnerabilities \cite{10.1145/3319008.3319029}.
Testing installed plugins for vulnerabilities can give a false sense of security, as Murphy et al. discovered: They examined 11 WordPress vulnerability scanner plugins and found that none identified all vulnerable plugins \cite{9509274}.

Furthermore, WordPress and its plugins have been used for evaluations in academic work, e.g.  \cite{tashenova2026sentinelcms,shezan2023chkplug,neef2024all}.
Niemietz et al. analyzed the top 100 Joomla-CMS plugins for vulnerabilities \cite{niemietz2021100bugsrowsecurity}, but do not consider exposed log files. 
Outside academia, large-scale vulnerability analysis of CMS extensions has been pursued as well. For example, wpgarlic\footnote{\url{https://github.com/kazet/wpgarlic/}, accessed: 2026-09-15} is a fuzzer for WordPress plugin that has uncovered a large number of vulnerabilities, although not directly supporting the identification of exposed logs. 

Therefore, the exposure of log files by Content Management Systems plugins remains to be studied, and we focus on WordPress as it is the CMS with the largest market share.

\subsection{Log File Exposure in WordPress Plugins}
WordPress is developed in PHP and, thus, consists of several \texttt{.php} files that are placed in a web server's \texttt{DocumentRoot}, which, by default, makes these files accessible. 
While correctly configured web servers will pass requests targeting \texttt{}{.php} files to the PHP engine for processing, other file extensions (e.g. \texttt{.txt} or \texttt{.log}) might be served by the web server directly.
However, in order for WordPress to provide its full functionality, such as uploading documents or images, or installing and updating plugins, some of its subfolders need to be writable by the web server. 

Plugin developers might want to provide their user-base with precise and helpful assistance should unexpected problems or incompatibilities with other plugins occur.
For these or other purposes, a plugin might want to write (debug) log files, which requires a writable folder.
As established earlier, the \texttt{wp-content/}, \texttt{uploads/}, or the plugin's own folder has a high chance to exist and be writable. 
Other folder locations might not be writable if the WordPress was hardened, e.g. using the official guide \cite{wordpressHardeningWordPress}. 
Thus, some log files end up in the web server's DocumentRoot and their contents may become accessible if not protected. 

Not all plugins succeed in sufficiently protecting the created log files, as Neef showed in an industry-study in 2020 \cite{detectifyWordpressPlugins}. 
A handful of popular plugins with over 1M active installations were exposing log files with sensitive information.
While not all exposures can be attributed to a plugin, their developer might have made assumptions about the WordPress instance's system or web server configuration.
For example, a \texttt{.htaccess} file can be used to block access to specific files or directories, but it is not supported by all web servers: Apache\footnote{\url{https://httpd.apache.org/}, accessed: 2026-09-15} supports it, while NGINX or others do not \cite{archiveNGINXLikeApache}, voiding this protection mechanism.
Another example is using a random or secret value in file names, but not preventing directory listing in the folder which exposes the file.

The log file contents are determined by each plugin and, thus, can range from irrelevant to sensitive. 
While error messages, stack traces, and directory paths can be helpful for attackers as additional information to prepare attacks, the exposure of usernames, credentials, or secrets could pose an immediate threat to the availability, integrity, or confidentiality of the website (e.g. \cite{easywpsmtp0day,sentineloneCVE202413513Oliver,managedwpMitigatingSensitive}).
More recently, a log file disclosure in the Post SMTP plugin put 400,000 websites at risk of admin account takeover in 2025 \cite{postsmtp0day}.

In general, such information disclosures fall into the \emph{A01:2025 - Broken Access Control} and \emph{A09:2025 - Security Logging and Alerting Failures} categories of the OWASP Top 10 of the most common security issues \cite{owaspOWASP102025}.
The Common Weakness Enumeration\footnote{\url{https://cwe.mitre.org/}, accessed: 2026-09-15} (CWE) tracks these as \emph{CWE-200: Exposure of Sensitive Information to an Unauthorized Actor}, \emph{CWE-215: Insertion of Sensitive Information Into Debugging Code}, and \emph{CWE-532: Insertion of Sensitive Information into Log File}.
If personal identifiable information (PII) is disclosed, \emph{CWE-359: Exposure of Private Personal Information to an Unauthorized Actor} becomes applicable.

According to the work of Mesa et al. \cite{10.1145/3233027.3233042}, information exposure is among the top 10 most common vulnerabilities in WordPress plugins, which further motivates this work and its agentic LLM-based approach to potentially identify more of such issues.

\subsection{Large Language Models for Plugin Vulnerability Detection}
Traditional static application security testing (SAST) tools such as Semgrep\footnote{\url{https://github.com/semgrep/semgrep-rules/tree/develop/php/wordpress-plugins/security/audit}, accessed: 2026-09-15}, Psalm\footnote{\url{https://psalm.dev/docs/}, accessed: 2026-09-14}, or RIPS\footnote{\url{https://github.com/robocoder/rips-scanner}, accessed: 2026-09-15} predate LLMs, but these do not currently come with detection patterns for exposed log files. 
Writing and defining the rules and patterns to detect exposures with these tools requires substantial engineering efforts and understanding of their capabilities. With LLMs and their reasoning capabilities, the bar is lowered to providing an analysis environment and a prompt.

Using Large Language Models (LLMs) for vulnerability detection (e.g. \cite{LU2024112031,guo2024outside,du2024vul}), validation (e.g. \cite{li2026execution,ghosh2025cve,nitin2025faultline}), explanation (e.g. \cite{kumari2026vuln2action,germano2025study}), or remediation (e.g. \cite{kulsum2024case,11396517}) is a trending and promising area of research at the time of writing.

Risse et al. argue that function-level security analysis does not provide enough context  to reliably determine a vulnerability \cite{risse2025}. 
Instead, repository-level access \cite{repoaudit,benchmarking-vuln-dect-repos} and the use of agents \cite{react} can lead to higher detection rates. 
Thus, our work also follows an agent-based analysis and adopts the repository-level context idea by providing a plugin's complete source code for analysis.

However, many of these publications focus on software application-related programming languages.
The usage of LLMs in the context of web applications, especially Content Management Systems such as WordPress, has not been widely studied yet, further motivating this work.

Cao et al. examined the use of LLMs for PHP-based web applications and improved the detection capabilities \cite{cao2024realvul}.
In a preliminary study, Ng et al. used WordPress as a target to discover that LLMs can be used to detect broken access control based on web paths \cite{10902771}.
Fang et al. showed that LLM agents can autonomously exploit 1-day web vulnerabilities based on CVEs, including WordPress \cite{fang2024llmagentsautonomouslyexploit}.
Afterwards, \emph{CVE Bench} was proposed as a benchmark to test LLM agents' abilities, which also features WordPress plugin vulnerabilities \cite{zhu2025cve}.
More recently, Leng et al. proposed LLM4Patch to help identify vulnerability patches in commit messages of WordPress plugins \cite{leng2025poster}.

Thus, our work aims to further close this gap by providing insights on how LLMs can help detect potentially severe security issues in CMS plugins.

\section{Methodology}\label{sec:methodology}
This section introduces our agentic analysis framework with which we systematically identify potential log file exposures in WordPress plugins. 
Our methodology comprises three major steps: plugin selection, agentic plugin analysis, and manual validation. 

\subsection{CMS and Plugin Selection}\label{sec:meth:pluginselection}
\begin{figure}[t]
    \centering
    \includegraphics[width=0.75\linewidth]{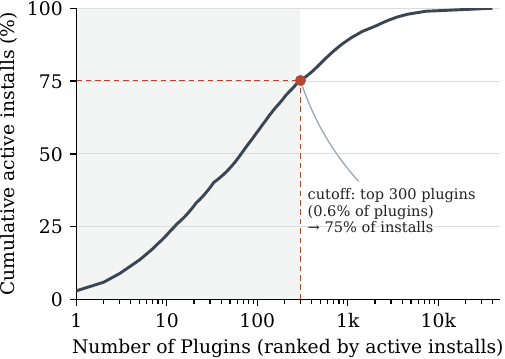}
    \caption{Active install count coverage of all unique WordPress plugins from the official plugin store (accessed: March 2026). The red marker depicts our selected top 300 plugins covering 75\% of the total active installation count.}
    \label{fig:pluginselection}
\end{figure}

Similar to related work (\Cref{sec:backgroundrelwork}), we focus on the WordPress ecosystem as a representative for other CMS and their comparable plugin systems, as it has almost 60\% market share and a gap of over 50\% to its nearest competitor.

The plugin selection for our analysis was influenced by the conclusion of Ruohonen \cite{10.1145/3319008.3319029} that more popular plugins can be of higher interest to attackers. 
Thus, we retrieved the full plugin catalog from WordPress' plugin system API\footnote{\url{https://api.wordpress.org/plugins/info/1.2/}, accessed: 2026-09-15} and sorted the unique plugins by the highest \emph{active installation} count. 
As shown in \Cref{fig:pluginselection}, we selected the top 300 plugins (0.6\% of all available plugins), covering over 75\% (250M) of the cumulative active installations,
providing broad real-world coverage while keeping the manual validation workload acceptable.
The most popular plugins in our selection have over 10M active installations each, with the lowest being over 100,000.
The full plugin list is available in our repository (see \Cref{sec:disc:openscience}).

\begin{figure*}[t]
    \centering
    \includegraphics[width=0.75\linewidth]{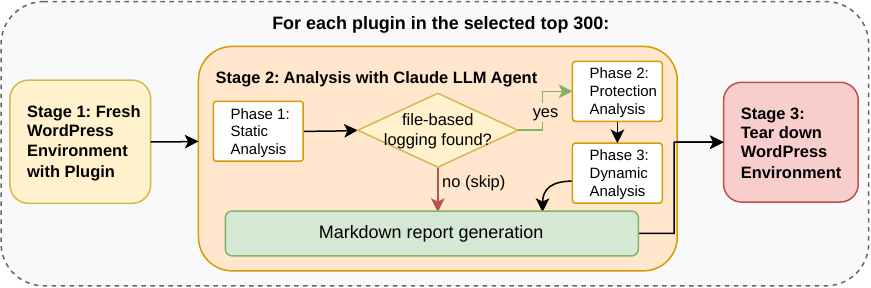}
    \caption{The three analysis stages of our agentic framework to detect potential log file exposures consisting of a setup and teardown stage, as well as an agentic analysis stage with static, protection, dynamic analysis phases.}
    \label{fig:meth:aianalysis}
\end{figure*}

\subsection{Agentic LLM-based Analysis Framework}\label{sec:meth:agenticanalysis}
As outlined in \Cref{sec:backgroundrelwork}, we decided to implement an agentic LLM-based framework for our potential log file exposure analysis of the top 300 selected WordPress plugins. 
All data and code will be open-sourced to foster future work and allow developers to test their own plugins as described in \Cref{sec:disc:openscience}.

For the LLM, we chose Anthropic's Claude Opus 4.6\footnote{\url{https://www.anthropic.com/news/claude-opus-4-6}, accessed: 2026-09-15}, as this was the most recent frontier and SOTA model available at the time of analysis, with its own \texttt{claude}-code harness\footnote{\url{https://claude.com/product/claude-code}, accessed: 2026-09-15} in \texttt{--dangerously- skip-permissions} mode and the \emph{20x MAX} subscription plan, which was cheaper than API-based usage (see \Cref{sec:dis:useofllms}). The main prompt is in \Cref{app:mainprompt} and the environment used WordPress (6.9.4) on Apache. All other technical details and data are included in our repository (\Cref{sec:disc:openscience}).

\Cref{fig:meth:aianalysis} visualizes the three implemented stages in our agentic framework.
Using a coordinator python script, we ran the agentic framework against each plugin. 
While stage 1 and stage 3 were responsible to set up and tear down a fresh analysis environment, stage 2 is where the LLM-based static and dynamic analysis happened in three phases.

\subsubsection{Stage 1 \& 3: Setup and Tear down of the Analysis Environment}
For each analysis, stage 1 and stage 3 created an ephemeral Apache-based WordPress instance inside a Docker container. 
Stage 1 started an initialization script that configured a standardized multi-site WordPress page\footnote{\url{https://github.com/Automattic/wpscan-vulnerability-test-bench}, accessed: 2026-09-15}, created a set of predefined accounts (one for each WordPress role), and installed the target plugin. 
After the stage 2 analysis, stage 3 was responsible to tear down all Docker containers, so that the next analysis would start over with a fresh and clean instance.

This design choice was deliberately made to ensure only \emph{one} plugin (and its required dependencies) was analyzed at once to prevent cross-plugin interference. 
Furthermore, it also ensured that potential log file exposures could be attributed to the tested plugin, and the agent could focus on a single plugin.

\subsubsection{Stage 2: Agentic Analysis}
For the agentic analysis, we provided claude-code a main prompt (see \Cref{lst:mainprompt} in Appendix \Cref{app:mainprompt}) stating the analysis objectives. 
The main prompt referenced two additional prompt files, \texttt{PROMPT.md} and \texttt{CLAUDE.md}, which provided further in-depth information about the evaluation environment, the analysis phases and how the markdown report was supposed to be generated. 
All prompts are available in our repository (see \Cref{sec:disc:openscience}).

\paragraph{Phase 1: Static Analysis}
The LLM-agent was instructed to extract the installed plugin's source code from the docker container for analysis.
Next, it should examine the source code for file-based logging using common file-writing code paths as guidance:
\pagebreak
\begin{itemize}
  \item \textbf{Direct file writes:}
        \texttt{file\_put\_contents}, \texttt{fopen}, \texttt{fwrite},
        \texttt{fputs}.
  \item \textbf{PHP error logging:}
        \texttt{error\_log} calls with an explicit file destination.
  \item \textbf{Logger abstractions:}
        method calls such as \texttt{->log()}, \texttt{->debug()},
        \texttt{->error()}; usage of Monolog or WooCommerce's \texttt{WC\_Logger} API, etc.
  \item \textbf{Path construction:}
        expressions involving \texttt{WP\_CONTENT\_DIR}, \texttt{ABSPATH},\\
        \texttt{wp\_upload\_dir()},\texttt{plugin\_dir\_path()},
        or \texttt{\_\_DIR\_\_}.
\end{itemize}
For each identified log-writing code path, it should trace the call chain and identify additional properties, such as the full file path, triggers for writing the file, the data that is written, and more.
We focus on \texttt{error\_log} calls with an explicit file destination, as PHP's error log resides outside the DocumentRoot by default. 
If no log file paths were identified, the next two phases were to be skipped.

\paragraph{Phase 2: Protection Analysis}
For each identified log path, the agent should determine if (and what) protections are implemented to prevent access to the log file. 
For example, it should check for \texttt{.htaccess}, \texttt{index.php}, \texttt{index.html}, randomized file paths, or other protective measures.

\paragraph{Phase 3: Dynamic Analysis}
In this phase, the agent was instructed to actively trigger the log file creation for each identified log file and observe the log file's creation in the running WordPress instance, as well as its accessibility over HTTP and the effectiveness of the implemented protections.
In particular, we provided the agent with pre-registered accounts with different roles and access to the web-interface of the WordPress instance as well as access to the WordPress container, so it could freely interact with the instance and validate log file exposures over HTTP or the file system.

\paragraph{Report Generation}
Finally, for each plugin the framework was instructed to generate a structured markdown report in the \texttt{findings/} folder. 
The coordinator script also created a copy of claude-code's commandline output in \texttt{output/}.
The markdown reports contain information about the LLM's analysis and findings about each potentially exposed log file. 
These reports were the basis for the manual validation and our results.

\subsection{Manual Validation}\label{sec:meth:validation}
Since LLMs are inherently prone to hallucinations, i.e. reporting non-existing or invalid issues, we deliberately decided to manually review all generated vulnerability reports and their log file path findings, instead of using another \emph{Judge} LLM or agent.
This means, that for each plugin and potential log file exposure, we took the LLM-produced report and instantiated the stage 1 analysis environment to reproduce and validate the log file exposure information. 

Based on the validated information, we populated an \texttt{analysis.csv} file tracking all exposure properties, such as log file paths, protection mechanisms, and comments about the findings validity, for each plugin and log file.

Additionally, we implemented a deliberately permissive AST-based scanner as a baseline to help detect false negatives, which we discuss in \Cref{sec:dis:useofllms}.  

\section{Results}\label{sec:results}

In this section, we first report the aggregate outcome of the analysis. 
We then introduce a taxonomy that structures the observed log file paths and protection mechanisms, and finally apply this taxonomy to quantify how the patterns are distributed and combined in practice.

\subsection{Log File Exposure Analysis}\label{sec:results:logfiles}

\begin{table}[t]
\centering
\caption{Overview of the analysis success and failure rates.}
\label{tab:counts}
\small
\begin{tabularx}{\linewidth}{@{}Xr@{}}
\toprule
\textbf{Metric} & \textbf{Count} \\
\midrule
Top WordPress plugins selected        & 300 \\
\midrule 
Agentic analysis: & \\
\quad Analysis failures (e.g. timeout)            & 13 \\
\quad Successfully analyzed     & 287 \\
\quad\quad Plugins without identified file logging     & 225 \\
\quad\quad Plugins with identified file logging       & 62 \\
\quad\quad Potential log file exposures identified     & 81 \\
\quad\quad\quad Manually successfully validated log files & 79 \\
\quad\quad\quad Log file findings with minor inaccuracies      & 2 \\
\midrule
AST-based analysis: & \\
\quad Identified plugins with potential file logging                     & 165 \\
\quad Also identified by agentic approach       & 58 \\
\quad Not identified by agentic approach     & 107 \\
\quad\quad Manually validated false positives                & 100 \\
\quad\quad Manually confirmed file logging instances       & 7 \\
\quad\quad\quad Timeout in agentic approach       & 6 \\
\quad\quad\quad True agentic based false negative & 1 \\
\midrule
 Not identified by AST, but by agentic approach & 4 \\
\bottomrule
\end{tabularx}
\end{table}

\Cref{tab:counts} contains the counts for the agentic analysis and our AST-powered baseline.
\subsubsection{Agentic Analysis}
Of the 300 plugins in our selection, 13 did not produce a finding (see \Cref{sec:disc:limitations}), leaving 287 analyzed plugins with generated findings.
For the vast majority (225 plugins, 78\%), no file-based logging code paths could be identified and, thus, no dynamic analysis was performed.
The remaining 22\% (62 plugins) created one or more log file, resulting in a total of 81 potential log file exposures. 

Our manual validation showed that the LLM-generated log file findings in the reports were valid in 79 cases, where valid means that we could reproduce and verify the information provided in the finding. 
Only in 2 cases, there were slight discrepancies, e.g. an \texttt{index.php} present that we could not confirm, or a different log location. 

The precision of our approach is 98\% with 79/81 identified log files manually validated, and a recall of 98\% (62/63) for the successfully analyzed plugins by our framework.

\subsubsection{AST-based Analysis}
For comparison, we built a permissive, non-LLM AST-parsing script to look for the logging guidance patterns (\Cref{sec:meth:agenticanalysis}) in the plugins' first-party source (excluding imported dependencies) as a baseline. While the script has no data-flow capabilities to determine where or how a log file is written, it flagged 165/300 plugins. 58 of these our framework identified as well, and the other 107 were manually reviewed: The script over-reported the majority due to normal file-write patterns, \texttt{error\_log} without a file destination, or other factors. In fact, only 7 plugins were manually confirmed to write log files. Six of these were not identified by our framework because the analysis \emph{timed out}, and 1 plugin (\emph{ninja-forms}), which used database-logging \emph{and} file-logging in one specific file, was missed by the farmework despite completing.
Conversely, the AST-baseline missed 4 plugins which the agentic approach discovered. 

Thus, over all known log-writing plugins, including the findings from the baseline, the recall of our framework drops to 90\% (62/69).


\begin{table*}[t]
\centering
\caption{Taxonomy of log file paths and protection mechanisms.} 
\label{tab:taxonomy}
\renewcommand{\arraystretch}{0.95}
\begin{tabularx}{\linewidth}{@{}l l X@{}}
\toprule
\textbf{Axis} & \textbf{Category} & \textbf{Description and security relevance} \\
\midrule
\multirow{6}{*}{\makecell[l]{\textbf{Path}\\(naming /\\location)}}
 & Static      & A hardcoded path; known to an attacker \\
 & Date-based  & Contains a date (e.g. YYYY-MM-DD); feasibly guessable \\
 & Hash-based  & Contains a hash value; unguessable \\
 & Secret-based      & Contains a hardly guessable value (e.g. random name) \\
 & \texttt{error\_log} & Logs to PHP's \texttt{error\_log} \\
 & Other       & Path feature not captured by the above \\
\midrule
\multirow{7}{*}{\makecell[l]{\textbf{Protection}\\(access /\\creation)}}
 & PHP stub        & e.g. \texttt{<?php exit; ?>} and \texttt{.php} suffix prevents output \\
 & Index file      & e.g. \texttt{index.php} or \texttt{index.html}; prevents directory listing \\
 & \texttt{.htaccess}       & Apache-specific configuration file with access control rules \\
 & Dir.-listing off & Directory listing disabled (via index file or .htaccess) \\ 
 & PHP constant    & Special PHP constant (e.g. \texttt{WP\_DEBUG}) needs to be defined \\
 & Manual change & Manual setting or code changes prior to log creation\\
 & Other    & A different protection mechanism, e.g. outside the DocumentRoot \\
& None & No protection mechanism \\
 \bottomrule
\end{tabularx}
\end{table*}
\subsection{Taxonomy of Log File Paths and Protection Mechanisms}

In order to create the basis for answering our research questions, we first distill our findings into a taxonomy on two axes: \emph{how the file is named and located} (path) and \emph{how it is shielded from unauthorized access} (protection). 
\Cref{tab:taxonomy} summarizes the resulting categories, which are an extension of  \cite{detectifyWordpressPlugins}. Our categories are not mutually exclusive.
For example, a log file can combine several path features and protection mechanisms, e.g. \texttt{\seqsplit{wp-content/uploads/plugin-<secret>-<YYYY-MM-DD>-log.php}}.

\paragraph{Path categories}
We group the path categories into two security-relevant classes. 
A log file path is \emph{predictable} when it is \emph{static} or \emph{date}-based, as the number of years, months or days are feasible to enumerate. 
It is \emph{obscured} when it contains a \emph{hash} or \emph{secret} value unknown to an attacker.
Therefore, we deem a log file \emph{guessable} only if \emph{predictable} and without \emph{obscured} features.

\paragraph{Protection categories}
The protection categories fall into two distinct groups. 
\emph{Access-control} mechanisms (PHP stub, index file, .htaccess, directory listing) try to limit access \emph{after} a log file was created.
\emph{Creation-control} mechanisms (PHP constant, manual change) define conditions \emph{before} a log file is created.
Thus, \emph{creation-control} prevents a log file from being created by default, but it does not necessarily protect it from being accessible once created. 
\begin{figure*}[t]
    \centering
    \begin{subfigure}{0.5\linewidth}
        \centering
        \includegraphics[width=\linewidth]{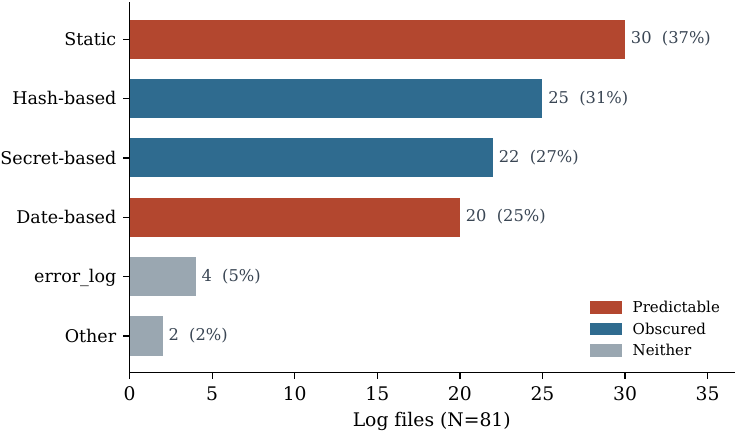}
        \caption{Prevalence of each path feature.}
    \end{subfigure}\hfill
    \begin{subfigure}{0.45\linewidth}
        \centering
        \includegraphics[width=\linewidth]{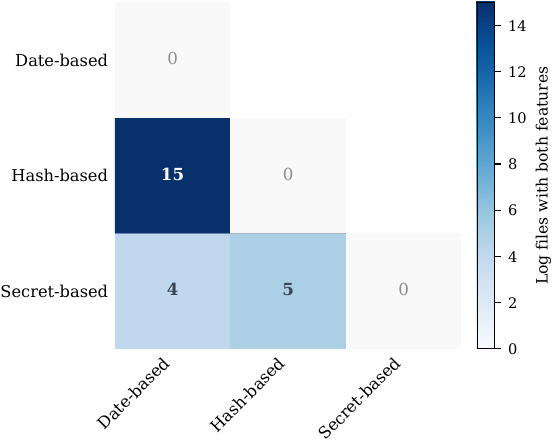}
        \caption{Pairwise co-occurrence of date-, hash-, secret-based path features.}
    \end{subfigure}
    \caption{Path feature distribution and co-occurrence across the 81 log files. In (b), only co-occuring features are shown as the remaining  only occur in isolation.}
    \label{fig:pathpatterns}
\end{figure*}

\pagebreak

\subsection{Identified Log File Exposure Patterns}\label{sec:res:patterns}
We now apply the taxonomy to the 81 potential log file exposures.

\subsubsection{Path patterns}
\Cref{fig:pathpatterns}(a) shows the \emph{per-feature} prevalences for the 81 log files.
\emph{Static} paths are most common (37\%), followed by \emph{hash}-based (31\%), \emph{secret}-based (27\%), and \emph{date}-based (25\%) features. \emph{PHP error\_log} occurs in four cases, and \emph{other} in two times.
In practice, a path can have multiple features that complement each other, so a path with a predictable feature is still not guessable if it also contain an obscurring feature. 
\Cref{fig:pathpatterns}(b) shows which path features co-occur. By definition, only \emph{hash}, \emph{secrets}, and \emph{date} can co-occur within a path. Only two log files have all three features, and except for 3 log files, the \emph{date} feature always co-occurs with a \emph{hash} or \emph{secret} feature in our dataset, and the most common combination is \emph{date}+\emph{hash}.

\begin{figure*}
    \centering
    \begin{subfigure}{0.47\linewidth}
        \centering
        \includegraphics[width=\linewidth]{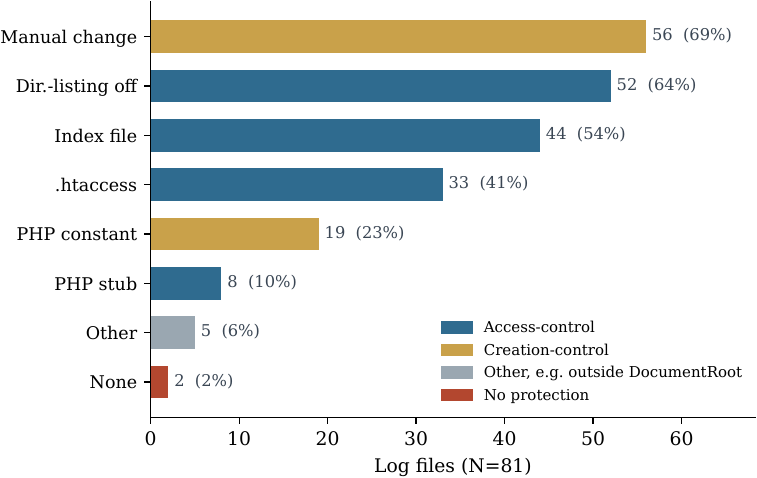}
        \caption{Prevalence of each protection mechanism.}
    \end{subfigure}\hfill
    \begin{subfigure}{0.53\linewidth}
        \centering
        \includegraphics[width=\linewidth]{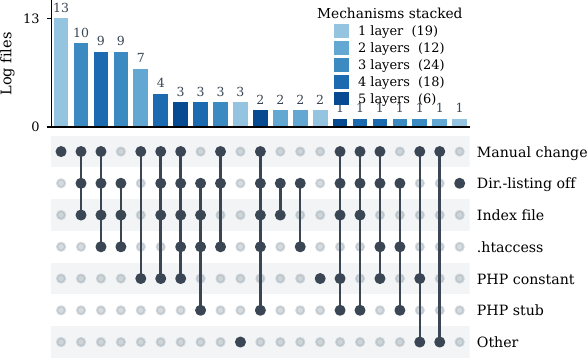}
        \caption{Number of stacked and co-occurring protection mechanisms.}
    \end{subfigure}
    \caption{Protection mechanism distribution and co-occurrence across all log files. The two log files with \emph{no protection} (0 layers) are omitted in (b).}
    \label{fig:protpatterns}
\end{figure*}

\subsubsection{Protection patterns}
\Cref{fig:protpatterns}(a) shows the prevalence of each protection mechanism. 
The most common protection is \emph{creation-control}, i.e. the log files are only created after a \emph{manual change} (69\%) or a \emph{PHP constant} is defined (23\%).
Disabling directory listing (64\%) is the most common \emph{access-control} mechanism, followed by \emph{index files} (54\%), and \emph{.htaccess} (41\%). The more robust \emph{PHP stub}s are only used in 8 cases. 
From the \emph{Others} category, 4 cases are PHP's default \texttt{error\_log()} and 1 log file whose name can be chosen by the user and the plugin ensuring it is not within the DocumentRoot.
Only two files have \emph{no protection}, but one file is written outside the DocumentRoot and the other to a user-supplied location (likely to be outside the DocumentRoot), thus remain unexposed.
\Cref{fig:protpatterns}(b) shows that \emph{access-control} features are typically stacked together. The most common number is 3 stacks (24 times), followed by four stacks (18 times). Only 6 log files are covered with 5 protection mechanisms, but 19 rely on only one.

\subsection{Strongest and Weakest Configurations}
The strongest configurations emerge when \emph{creation-control} and \emph{access-control} are both present and cover many categories. \Cref{fig:protpatterns}(b) shows that using multiple protection mechanisms is the norm.
On the other hand, 13 log files rely on the single \emph{manual change} protection, which can be sufficient depending on the change required (e.g. code-changes are a higher barrier than toggling a UI switch). However, some code-changes can be performed by administrators through the built-in code-editor.

\begin{figure}[t]
    \centering
    \includegraphics[width=\linewidth]{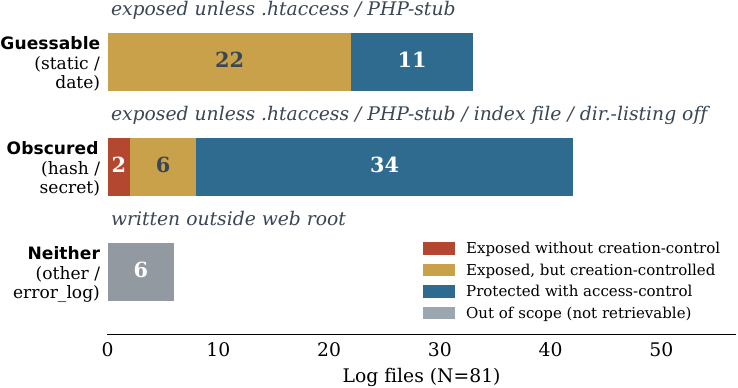}
    \caption{The relation between path features and protection mechanisms.}
    \label{fig:exposure}
\end{figure}

\Cref{fig:exposure} shows the relationship between path features and relevant protection mechanisms. 
A \emph{guessable} path must, by definition, be assumed to be known to an attacker, but access can be prevented with an \texttt{.htaccess} deny rule or a PHP stub. Disabled directory listing will not protect such files. 
On the other hand, a properly \emph{obscured} path can become accessible (given no other access control) if directory listing is enabled, thereby revealing the unknown values.

From this perspective, 22 of the 33 guessable files and 8 of the 42 obscured files are at risk of exposure, while the remaining files are shielded by access controls.
However, taking the \emph{creation-control} into account for the 22 files, it actually brings the number down to zero, as no log file is exposed by default.
Only two of the 8 obscured log files have no \emph{creation-control} and might become exposed, if their filenames are revealed.

\section{Discussion}\label{sec:discussion}
In this section we discuss our results with regard to our research questions, limitations, future work, and ethical considerations.

\subsection{RQ1: Prevalence of Log File Exposures in WordPress Plugins}
Our results show that file-based logging is common in the top 300 plugins of the most popular CMS (WordPress), and several potential log file exposures were identified among our framework's findings, but none are directly at risk of exploitation. 
Only roughly every fifth (62 of 287) successfully analyzed plugin created at least one log file (81 in total), but these are guarded by one or several protective mechanisms.
Although not directly comparable due to methodological and dataset differences, our results appear different from the industry study in 2020 \cite{detectifyWordpressPlugins}, where multiple unprotected log file exposures were identified.
One explanation could be that awareness of log file exposures and defense-in-depth measures could have increased among developers of the top plugins over the last six years, but more research is needed and left to future work.

\subsection{RQ2: Log File Path and Protection Patterns}
To our surprise, the most common safeguard is not \emph{access-control}, but \emph{creation-control}. 69\% of the 81 log files (\Cref{sec:res:patterns}) are only created when the plugin-user manually enables related settings (e.g. ``enable debug log'' toggle) in the UI, or edits the code to enable logging functionality (e.g. by defining a constant such as \texttt{WP\_DEBUG} or \texttt{WP\_DEBUG\_LOG}). For 19 of the 81 log files (23\%) the required change is defining a special PHP constant. Thus, on a default installation without deliberate changes, these log files will never be created and, thus, are likely only used by developers.
It also explains why for the 22 \emph{guessable} log files without sufficient protective \emph{access-control} mechanisms, none are immediately at risk of exploitation.

However, whether a log file can be accessed by unauthorized attackers also depends on the server's configuration. If directory listing is globally enabled in the server's configuration, \emph{obscured} log file paths become discoverable and accessible if no additional \emph{access-control} is in place. 
Similarly, the \texttt{index} file names can be changed globally in the web server configuration, breaking the assumption about \texttt{index.php} or \texttt{index.html}.
Also, if the web server does not support \texttt{.htaccess} configuration files, e.g. NGINX, then that \emph{access-control} protection becomes ineffective for 41\% of the observed log files. 
Also, web administrators might choose to disallow PHP-execution of files in folders with user-supplied content, e.g. \texttt{wp-content/uploads/}, to mitigate attacks (e.g. remote code execution) which would void the \emph{PHP-stub} protection.
Therefore, we believe that some plugin developers make assumptions about the web server or its configuration (i.e. Apache with \texttt{.htaccess} support) on which their plugin gets installed.

For end-users with varying degrees of technical knowledge, to whom WordPress caters, it might not be obvious \emph{if}, \emph{when} or \emph{how} a plugin creates log files, \emph{what} is being logged, and \emph{how} these files are being protected. This creates a risk to end-users that an installation or usage of a plugin might accidentally put their website at risk of compromise, as the past has shown (\cite{easywpsmtp0day,postsmtp0day}).

\subsection{RQ3: Best Practices for Secure Log Files in CMS plugins}
From our observations we aim to derive the following best practices for developers of plugins and CMS in handling log files.

\subsubsection{Avoid log files in the DocumentRoot}
The most robust mitigation against log file exposures is not creating them in the web-accessible DocumentRoot in the first place. Temporary log files used for debugging could be written to ephemeral and non-accessible storage outside the DocumentRoot, such as \texttt{/tmp/} on Linux servers. 
Alternatively, as some of our analyzed plugins already offer, the log contents should be written to the CMS' database to avoid disclosure.

\subsubsection{Layered Defense-in-depth}
If file-based logging is unavoidable, it is crucial to implement multiple protective mechanisms. We recommend combining all of the following to cover several failure modes:
\begin{itemize}
    \item A \textbf{PHP-stub} and \texttt{.php}-suffix to prevent disclosure on direct access. 
    \item An \textbf{index} file to suppress directory listing in the log file's folder.
    \item An \textbf{.htaccess} file to deny access on Apache-based web servers. 
    \item An \textbf{obscured} file path with sufficiently random or secret values so that the URL cannot be guessed.
\end{itemize}

Also, logging should be made visible to the end-users to make them aware of the potential risk of exposed log files. The plugins should provide appropriate UI views where all created log files are being listed with an option for deletion.

\subsubsection{CMS-level Logging API}
Finally, CMS should implement and offer appropriate logging APIs that will allow plugin developers to create log files managed and secured by the CMS itself. 
That way, developers would not need to implement log file management themselves, which should eliminate the risk of log file exposure. Also, end-users could benefit by having all logs in a central and reviewable place. 

\subsection{Use of LLM-Agents for Log File Exposure Detection}\label{sec:dis:useofllms}
Our choice to use LLMs for this analysis was motivated by several factors. First, log file exposures can become the result of complex code paths (e.g. final paths assembled from constants, wrapper functions, or from options read from the database), potentially requiring complex detection rules for SAST tools. At the time of writing, SAST tools such as Semgrep or Psalm did not implement rules for log file exposure detection, excluding them from a comparison. Second, a log file write does not automatically equal exposure, as the different protection mechanisms can be implemented which need to be evaluated at runtime (e.g. \texttt{.htaccess}), thus requiring dynamic analysis capabilities. With its tool-calling capabilities and the WordPress environment, the LLM was able to interact with the instance to verify the findings. Lastly, LLMs allow for natural language descriptions of the goal, rather than implementing and maintaining SAST-engine specific rules for each logging idiom (e.g. direct writes, \texttt{WC\_Logger}, custom loggers, etc.). Nonetheless, we believe another promising approach could be the use of LLMs to generate such rules, or use SAST-tools in combination with agentic workflows, which we leave to future work. 

We found the generated reports to be precise, as we were able to reproduce 79 of 81 reported log file findings during manual validation (see \Cref{sec:meth:validation}).
The descriptions of the relevant code-paths and triggers of the log file creation were deemed very helpful. In fact, it greatly facilitated and accelerated the code-review and reproduction of the finding.

Giving the LLM-agent full access to each plugin's source code allowed it to analyze and trace through the execution flow to understand complex code-paths. 
Combined with the multi-phase static and dynamic analysis conducted by the LLM-agent, we believe it helped to verify and produce more accurate findings. 
For comparison, we built a permissive, non-LLM AST-parsing script to look for the logging guidance patterns (\Cref{sec:meth:agenticanalysis}) in the plugins' first-party source (excluding imported dependencies) as a baseline. While the script has no data-flow capabilities to determine where or how a log file is written, it still identified one plugin not identified by our framework (\Cref{sec:results:logfiles}). 
However, the baseline missed 4 plugins harder-to-model file-writing methods that our framework correctly identified. Thus, the agentic approach can reduce the false positives generated by a simple SAST script and identify more complex issues.  

The analysis cost $\sim$\$258 in equivalent Claude API usage (mean \$0.91, median \$0.69 per plugin). It took $\sim$  1800 minutes and about 300M tokens to complete. For each plugin, the analysis took a median of 5 minutes (mean 5.6, max 13.2) over a median of 18 agent reasoning steps.

\subsection{Limitations and Future Work}\label{sec:disc:limitations}
Our study has several limitations that also point to future work.

First, we limited our analysis to the top 300 WordPress plugins, which cover 75\% of the active plugin installations from the official WordPress plugin system. Future work could extend the analysis to the long tail of less popular plugins to determine if and how the prevalence of log file exposures and their mitigations change. 
Although WordPress has the biggest CMS market share by a large margin, a comparison to other CMS and their plugins could reveal further insights.

While our approach identified several potential log file exposures, which we manually validated, there remains a chance that the LLM-based analysis did not identify all log exposure instances, e.g. our baseline-script or agentic framework missed. In particular, the true recall could be lower if both missed log file exposures. However, we deliberately decided against including plugins with known file-exposure issues as a groundtruth, since these could be within a SOTA model's training dataset, biasing the results. Therefore, we encourage future work to repeat our study, e.g. with different frontier models, agentic setups, CMS or plugins.

Additionally, we provided the LLM one main prompt with analysis instructions and two additional file references for guidance, but did not monitor how closely the agent followed these instructions. Future work could review our captured outputs from the LLM-agent to learn how the instructions or LLM-based analysis could be improved, e.g. with other prompt designs or different/no guidance.
We acknowledge that LLMs are non-deterministic and single analysis-runs can lead to false-negatives, so we encourage future work to repeat our study with multiple iterations per plugin to gain run-to-run consistency. 
Furthermore, an ablation study would provide insights what effect the static or dynamic analysis phases of our framework have on the overall results. 

Of the 300 analyzed plugins, only 16 exceeded the 900 seconds analysis time limit, but 3 still produced a findings report. \Cref{tab:timeouts} in the Appendix lists these 13 plugins with their version, lines of code and number of files. While plugin and code complexity can be one reason to cause a timeout, network and initialization delays of the dynamic analysis environment (Phase 3) were counted against the execution time as well. Future work could re-run the analysis with raised timeouts.

We evaluated our log file exposure framework only against the traditional Apache+WordPress combination in its default, multisite configuration. While Apache is recommended by WordPress \cite{wordpressWordPressServers}, other web servers can be used as well, where certain protection mechanisms might apply (e.g. \texttt{.htaccess} not supported by NGINX). 
Future work could re-evaluate the risk assessments (e.g. \Cref{fig:exposure}) and log file exposures for CMS installations on non-Apache web servers.

A large-scale measurement could further establish the prevalence of the WordPress+web-server combinations actually used on the internet to determine whether shifts in assumptions and configurations are necessary.
Additionally, such a large-scale internet scan would allow to identify and quantify the exposures caused by web server (mis)configurations and help affected parties to remediate these by notifying them.

An analysis of the exposed log files' contents was deemed out of scope, as it would require deeper understanding on how a plugin is used and what data is processed. However, the type of exposed data determines the exposure's severity (e.g. from benign debug lines to highly sensitive access tokens or credentials). Future work could perform such a classification or, after approval by an IRB, conduct a large-scale study of live exposed files to rate the severity of such issues.

As stated earlier, we did not compare our approach against static analysis tools such as Semgrep or Psalm, as no suitable rules for log file exposures are publicly available. Similarly, we did not compare against community projects like WPScan or Nuclei as these check for specific publicly known log file exposures, which are likely present in the LLM's training data, which would bias the results. To the best of our knowledge, these tools cannot be used to uncover previously unknown log file exposures, which was the goal of analyzing the most recent versions of the top WordPress plugins in this work.  

\subsection{Open Science and Ethical Considerations}\label{sec:disc:openscience}
To foster reproducibility and to allow developers to audit their own plugins, we open-source the complete analysis framework and all related code in the following repository upon acceptance: \url{https://github.com/gehaxelt/Agentic-Detection-of-Potential-Log-File-Exposures-in-CMS}.
The plugin analysis data will be released after closely reviewing the findings again and following coordinated vulnerability disclosure (CVD) where necessary, and providing developers a sensible amount of time to implement additional protections. 

From an ethics perspective, this work was conducted with a defensive intent and did not require approval by the authors' universities. 
Furthermore, all analysis was performed against local instances and we did not try to access or collect exposed log files from live third-party websites. 
We judged the identified issues on their severity and immediate risk to end-users and could not identify immediate high-risk disclosures, or otherwise we would have followed CVD practices as stated before. 
We believe that open-sourcing this framework poses a greater advantage for defense by allowing developers to detect and mitigate such issues to release a more secure plugin, rather than malicious actors using it with bad intentions. 

\section{Conclusion}\label{sec:conclusion}
We studied the potential exposure of log files created by third-party CMS plugins using our agentic LLM-based detection framework against the top 300 WordPress plugins that cover 75\% of active installations in the official plugin ecosystem. 
With our framework's static and dynamic analysis capabilities, we discovered 62 plugins writing 81 log files, which we manually reproduced and validated in 79 cases. Although we observed many protective measures and no directly exposed log files among our framework's findings, we find the the risk of exposure to be still just one misconfiguration away in many cases (RQ1). 

Furthermore, based on our data, we derived and extended a taxonomy of log file path and protection patterns and learned that the prominent safeguard is \emph{creation-control} instead of \emph{access-control} (RQ2). In fact, 69\% of the log files are only created after manual changes to the settings or code by the plugin user. But not all log files come with sufficient \emph{access-controls} once they are created, so an exposure risk remains that is not fully transparent to plugin users.

We noticed several protective patterns hinge on assumptions by developers about a web server's configuration to which a plugin is installed, which do not always hold. We have therefore compiled recommendations for developers to use multiple layers for defense-in-depth (RQ3).

Finally, we corroborate related work in that agentic, LLM-based security analysis frameworks are a useful tool to help identify potential log file exposures and accelerate the discovery and reproduction of security issues.

\section{Generative AI Use}
Claude Opus 4.8 was used for grammar and spelling checks when writing the paper, and for speeding up development of the evaluation and analysis scripts, but all AI-produced code was reviewed by the author.

\bibliographystyle{ACM-Reference-Format}
\bibliography{bibliography}

\appendix
\section{Appendix}

\subsection{Main Prompt}\label{app:mainprompt}
\Cref{lst:mainprompt} shows the main prompt given to the LLM agent with references to two files with additional guidance. 
\begin{lstlisting}[style=markdown,language=python, caption={Main prompt given to the claude agent.},label={lst:mainprompt}]
def build_prompt(plugin, findings_dir_name):
    name = html.unescape(plugin["name"])
    slug = plugin["slug"]
    version = plugin["version"]
    active_installs = plugin["active_installs"]

    return f"""You are analyzing WordPress plugin "{name}" (slug: `{slug}`, version: `{version}`, {active_installs:,} active installs) for publicly accessible log file exposure.

Follow the methodology in PROMPT.md and CLAUDE.md exactly. This is scientific research - be thorough.

Steps:
1. Create the plugin config: config/{slug}.json
2. Start the environment: ./scripts/run.sh --config {slug}.json
3. Wait for readiness, handle any dependency issues (e.g. WooCommerce requirement)
4. Extract the plugin source and perform THOROUGH static analysis - read and understand the code, don't just grep
5. If file-based logging is found, perform dynamic verification (Phase 2 + Phase 3)
6. Write the finding to {findings_dir_name}/{slug}-{version}.md using the exact format from PROMPT.md
7. Teardown: ./scripts/teardown.sh

Important:
- The finding MUST be written to {findings_dir_name}/{slug}-{version}.md before you finish (this path overrides any `findings/` reference in CLAUDE.md or PROMPT.md)
- If the plugin has no file-based logging, still write a finding stating that
- Do NOT skip any phase. Do NOT take shortcuts. Read the actual source code.
- Follow through wrapper functions and helper classes to trace actual file paths
- Check for ALL file-writing patterns: file_put_contents, fopen, fwrite, fputs, error_log with file dest, custom logger classes, Monolog, WC_Logger, etc.
- Check path construction: WP_CONTENT_DIR, ABSPATH, plugin_dir_path, wp_upload_dir, __DIR__, dirname(__FILE__)
- Check for log-related constants, options, and configurable paths
"""
\end{lstlisting}

\subsection{Timed-Out Plugins}\label{app:timeoutedplugins}
\Cref{tab:timeouts} shows the 13 plugins with their version, install count, lines of code and number of files (determined with \texttt{cloc}\footnote{\url{https://github.com/aldanial/cloc}, accessed: 2026-09-15}) whose analysis exceeded the
900s time limit without producing a finding (\Cref{sec:disc:limitations}).
\begin{table}[h]
\centering
\caption{The 13 timed-out plugins without a finding.}
\label{tab:timeouts}
\small
\setlength{\tabcolsep}{5pt}
\begin{tabular}{@{}llrrr@{}}
\toprule
\textbf{Plugin (slug)} & \textbf{Version} & \textbf{Installs} & \textbf{Files} & \textbf{LOC} \\
\midrule
\texttt{google-listings-and-ads}     & 3.6.0  & 900K\,+ & 4,618 & 360,808 \\
\texttt{wp-optimize}                 & 4.5.1  & 1M\,+   &   894 & 117,796 \\
\texttt{facebook-for-woocommerce}    & 3.6.0  & 500K\,+ &   567 &  61,061 \\
\texttt{fast-indexing-api}           & 1.1.22 & 200K\,+ &   688 &  58,844 \\
\texttt{mailchimp-for-woocommerce}   & 6.0    & 300K\,+ &   179 &  55,537 \\
\texttt{woocommerce-paypal-payments} & 3.4.1  & 800K\,+ &   978 &  54,338 \\
\texttt{woocommerce-services}        & 3.5.1  & 600K\,+ &   472 &  33,621 \\
\texttt{userfeedback-lite}           & 1.11.1 & 200K\,+ & 2,309 &  27,732 \\
\texttt{pinterest-for-woocommerce}   & 1.4.25 & 300K\,+ &   364 &  24,310 \\
\texttt{woo-variation-swatches}      & 2.2.3  & 300K\,+ &    48 &  11,040 \\
\texttt{host-webfonts-local}         & 6.2.0  & 300K\,+ &    71 &   7,042 \\
\texttt{post-duplicator}             & 3.0.13 & 200K\,+ &    35 &   4,081 \\
\texttt{connect-polylang-elementor}  & 2.5.5  & 100K\,+ &   272 &   3,149 \\
\bottomrule
\end{tabular}
\end{table}

\end{document}